\documentclass[11pt]{article}
\usepackage{axodraw}
\usepackage{epsfig}
\usepackage{amsfonts}
\usepackage{bbm}
\input pix.sty

\newcommand{\mD}{m_\rmii{D}}
\newcommand{\mG}{m_\rmi{G}}

\newcommand{\Nc}{N_{\rm c}}

\newcommand{\rmO}{{\mathcal{O}}}

\def\lsi{\raise0.3ex\hbox{$<$\kern-0.75em\raise-1.1ex\hbox{$\sim$}}}
\def\gsi{\raise0.3ex\hbox{$>$\kern-0.75em\raise-1.1ex\hbox{$\sim$}}}
\newcommand{\lsim}{\mathop{\lsi}}
\newcommand{\gsim}{\mathop{\gsi}}

\newcommand{\nB}{n_\rmi{B}}
\newcommand{\rmii}[1]{{\mbox{\tiny\rm{#1}}}}
\newcommand{\re}{\mathop{\mbox{Re}}}
\newcommand{\im}{\mathop{\mbox{Im}}}

\newcommand{\Tint}[1]{{\hbox{$\sum$}\!\!\!\!\!\!\!\int\,}_{\!\!\!\!\raise-0.9ex\hbox{$\scriptstyle{#1}$}}}

\newcommand{\unit}{{\mathbbm{1}}} 
\newcommand{\bi}{\begin{itemize}}
\newcommand{\ei}{\end{itemize}}

\newcommand{\hide}[1]{ }
\def\ring{\mathaccent"7017} 
\def\ContA{\piccc{%
 \Lqu(5,20)(35,20)%
 \Lqu(55,20)(85,20)%
 \Lsc(5,10)(35,10)%
 \Lsc(55,10)(85,10)%
 \Asc(35,15)(5,270,90)%
 \Asc(55,15)(5,90,270)%
 \SetWidth{0.5} 
 \Line(0,15)(90,15)%
 \Line(45,17)(45,13)
}}
\def\ContB{\piccc{%
 \Lqu(5,20)(85,20)%
 \Laqu(5,10)(85,10)%
 \SetWidth{0.5} 
 \Line(0,15)(90,15)%
 \Line(45,17)(45,13)
}}
\def\ContC{\piccc{%
 \Aqu(45,15)(5,0,180)%
 \Asc(45,15)(5,180,360)%
 \SetWidth{0.5} 
 \Line(0,15)(90,15)%
 \Line(45,17)(45,13)
}}

\makeatletter \@addtoreset{equation}{section} \makeatother
\renewcommand{\theequation}{\arabic{section}.\arabic{equation}}
\makeatletter
\renewcommand\section{\@startsection {section}{1}{\z@}%
                                   {-5.5ex \@plus -1ex \@minus -.2ex}
                                   {2.3ex \@plus.2ex}%
                                   {\normalfont\large\bfseries}}
\renewcommand\subsection{\@startsection{subsection}{2}{\z@}%
                                     {-3.25ex\@plus -1ex \@minus -.2ex}%
                                     {1.5ex \@plus .2ex}%
                                     {\normalfont\normalsize\bfseries}}
\renewcommand\thesection {\@arabic\c@section}
\renewcommand\thesubsection   {\thesection.\@arabic\c@subsection}
\renewcommand{\@seccntformat}[1]{%
\csname the#1\endcsname.\hspace{1.0em}}
\makeatother

\begin{document}

\begin{titlepage}
\begin{flushright}
BI-TP 2007/14\\
MS-TP-07-16\\
arXiv:0707.2458\\ \vspace*{1cm}
\end{flushright}
\begin{centering}
\vfill

{\Large{\bf 
Thermal imaginary part of a real-time static potential \\[2mm] 
from classical lattice gauge theory simulations
}} 

\vspace{0.8cm}

M.~Laine$^\rmi{a}$, 
O.~Philipsen$^\rmi{b}$, 
M.~Tassler$^\rmi{b}$ 

\vspace{0.8cm}

$^\rmi{a}${\em
Faculty of Physics, University of Bielefeld, 
D-33501 Bielefeld, Germany\\}

\vspace*{0.3cm}
 
$^\rmi{b}${\em
Institute for Theoretical Physics, University of M\"unster, 
D-48149 M\"unster, Germany\\}

\vspace*{0.8cm}

\mbox{\bf Abstract}
 
\end{centering}

\vspace*{0.3cm}
 
\noindent
Recently, a finite-temperature real-time static potential has been
introduced via a \linebreak Schr\"odinger-type equation satisfied by 
a certain heavy quarkonium Green's function. Furthermore, it has been 
pointed out that it possesses an imaginary part, which induces a finite width 
for the tip of the quarkonium peak in the thermal dilepton production rate. 
The imaginary part originates from Landau-damping of low-frequency 
gauge fields, which are essentially classical due to their high occupation 
number. Here we show how the imaginary part can be measured with classical 
lattice gauge theory simulations, accounting non-perturbatively for the 
infrared sector of finite-temperature field theory. We demonstrate that 
a non-vanishing imaginary part indeed exists non-perturbatively; 
and that its value agrees semi-quantitatively with that predicted 
by Hard Loop resummed perturbation theory.

\vfill

 
\vspace*{1cm}
  
\noindent
September 2007

\vfill

\end{titlepage}

%
\section{Introduction}

The notion of a static potential, generalizing the potential that 
appears in the Schr\"odinger-equation of non-relativistic quantum 
mechanics, is thought to play a role for heavy quarkonium physics 
in QCD. If the energy of a two-quark system, $E$, is close to twice 
the mass of the heavy quark, $M$, so that the combined 
``kinetic energy'' of the two quarks, $E - 2 M$, is small compared 
with $M$, then we may assume the quarks to be ``static''
to a good approximation, moving 
only slowly in the attractive potential generated by the colour fields. 

To turn this intuitive picture into a quantitative description 
requires the use of effective field theory methods. At zero temperature 
various energy and momentum scales can be identified, the small expansion
parameter being related to the ratio $(E - 2 M)/M$. The relevant 
effective theory is called NRQCD~\cite{nrqcd0}, or one of 
its variants, like pNRQCD~\cite{pnrqcd};  
for reviews on the various effective theories
used for describing heavy quarkonium, see refs.~\cite{mb,nrqcd}. 
The static potential plays the role of a certain matching coefficient 
in these effective theories: it is related to, but not identical with, 
the non-perturbative static potential that is traditionally defined from 
a large Euclidean Wilson loop in lattice QCD.  

At finite temperatures, the situation becomes more complicated
than at zero temperature. Indeed, finite-temperature field theory 
possesses many momentum and energy scales of its own: gluonic momenta 
could parametrically be $k \sim \pi T, gT, g^2T$~\cite{linde,gpy}, 
while gluonic frequencies (energies) 
can be even softer, down to $E \sim g^4 T$~\cite{asy,dblog}. 
Here $T$ is the temperature 
and $g$ is the QCD gauge coupling. The relevant effective description
now depends on the relation of these scales to the scales already
appearing in the zero-temperature situation. 

In fact, at finite temperatures, the situation is quite complicated
even at the leading non-trivial order in $g$. This might be anticipated
from the fact alone that the definition of a static potential based on the 
Euclidean Wilson loop appears to lose its meaning: 
the Euclidean time direction becomes 
compact, and large Wilson loops do not possess the same interpretation 
as at zero temperature. Replacing the Wilson loop by a correlator
of Polyakov loops does not remedy the situation~\cite{MDNLO,jp}.
Moreover, physics lives in Minkowski spacetime, 
which at finite temperatures in general requires
a non-trivial analytic continuation~\cite{acont}.  

Recently an attempt was made to give
a proper definition of a static potential in this situation, 
in the sense of obtaining an object which has a direct connection
to the spectral function of the heavy quarkonium system
(at least up to some order in perturbation theory)~\cite{static}. Formally, 
the static potential could be defined as a certain coefficient in 
the large-$M$ expansion of an equation of motion satisfied by 
a suitable heavy quarkonium Green's function. At leading non-trivial
order, the corresponding object was computed in Hard Thermal 
Loop~\cite{htl} resummed perturbation theory in ref.~\cite{static}. 
It was found that, at least to this order, the static potential
{\em can} also be obtained from a specific analytic continuation
of the Wilson loop defined in Euclidean spacetime with a compact
time direction. At the same
time, this analytic continuation yields properties that are not 
familiar from the zero-temperature context: in particular, the
potential develops an imaginary part. 

It is the purpose of the present paper to elaborate on the existence
of an imaginary part. We start, in \se\ref{se:def}, by reviewing 
the definition(s) introduced in ref.~\cite{static}. In \se\ref{se:clgt}
we argue that the imaginary part of the static potential remains
non-zero in the classical limit, by computing it perturbatively in classical
lattice gauge theory. Given that perturbative computations at finite
temperatures may ultimately suffer from infrared divergences, 
we carry out non-perturbative Monte Carlo simulations in classical 
lattice gauge theory in \se\ref{se:sim}, and compare the results
with those of the perturbative computation. 
We conclude in \se\ref{se:concl}.

%
\section{Definition of a real-time static potential}
\la{se:def}

We start by defining a certain Green's function in hot QCD. 
Let $\vec{r}$ be a point-splitting vector, and $\hat\psi$ a generic
heavy quark field operator in the Heisenberg picture. Then we introduce
\be
 \check C_{>}(t,\vec{r}) 
 \equiv 
 \int \! {\rm d}^3 \vec{x}\,
 \Bigl\langle
  \hat{\!\bar\psi}\,\Bigl(t,\vec{x}+\frac{\vec{r}}{2}\Bigr)
  \gamma^\mu
  \, W 
  \, 
  \hat \psi\Bigl(t,\vec{x}-\frac{\vec{r}}{2}\Bigr) \;\; 
  \hat{\!\bar\psi}\,(0,\vec{0})
  \gamma_\mu
  \hat{\psi}(0,\vec{0})
 \Bigr\rangle
 \;, \la{poinsplit}
\ee
where $W$ is a Wilson line along a straight path connecting
the adjacent operators, 
inserted in order to keep the Green's function gauge-invariant;
the metric is $\eta_{\mu\nu} = \mathop{\mbox{diag}}$($+$$-$$-$$-$); 
and the expectation value refers to 
$\langle...\rangle\equiv \mathcal{Z}^{-1} \tr [\exp(-\hat H/T)(...)]$, 
where $\mathcal{Z}$ is the partition function, 
$\hat H$ is the QCD Hamiltonian operator, and $T$ is the temperature.

The significance of the Green's function in \eq\nr{poinsplit} is
that if we take the limit $\vec{r}\to\vec{0}$, and subsequently 
Fourier transform with respect to the time $t$, then we obtain
a function which is trivially related to the heavy quarkonium 
spectral function, $\rho(\omega)$, in the vector channel:
\be
 \rho(\omega) = \fr12 \Bigl( 1 - e^{-\frac{\omega}{T }}\Bigr)
 \int_{-\infty}^{\infty} \! {\rm d} t \, e^{i \omega t}
 \check C_{>}(t,\vec{0})
 \;.
\ee
On the other hand, keeping $\vec{r}\neq \vec{0}$ for the moment,
makes it easier to analyse this Green's function
in perturbation theory.  

Let us consider $\check C_{>}(t,\vec{r})$ in the 
limit that the heavy quark mass $M$ is very large. 
Then $\check C_{>}(t,\vec{r})$ satisfies 
a Schr\"odinger equation of the type
\be
 \biggl\{ i \partial_t - \biggl[ 2 M 
 + V_{>}(t,r)
 - \frac{\nabla_\vec{r}^2}{M}
 + \rmO\biggl(\frac{1}{M^2} \biggr)
 \biggr] \biggr\}  \check C_{>}(t,\vec{r}) = 0 
 \;,  \la{Seq}
\ee
with the initial condition
\be
 \check C_{>}(0,\vec{r}) = - 6 \Nc\, \delta^{(3)}(\vec{r})
 + \rmO\biggl(\frac{1}{M} \biggr)
 \;. \la{In0}
\ee
The terms shown explicitly in \eqs\nr{Seq}, \nr{In0} result from a tree-level 
computation; they also develop multiplicative radiative corrections
which we have omitted for simplicity. In contrast, 
the potential denoted by $V_{>}(t,r)$ originates only at 1-loop order. 
It can be defined as the coefficient scaling as $\rmO(M^0)$, 
after acting on $\check C_{>}(t,\vec{r})$ with the time 
derivative $i\partial_t$.

Now, as \eq\nr{Seq} shows, $V_{>}(t,r)$ can even be defined
in the limit $M\to\infty$, provided that the trivial factor $2 M$ is 
shifted away by a redefinition of time, as is standard in NRQCD. 
In ref.~\cite{static}, the computation in this limit was carried out
to 1-loop order,  $\rmO(g^2)$, in Hard Thermal Loop resummed 
perturbation theory~\cite{htl}. It was found that at this order
$V_{>}(t,r)$ can in fact be extracted from the equation 
\be
 i \partial_t C_{E}(it,r) \equiv V_{>}(t,r) C_{E}(it,r)
 \;,  \la{Seq2}
\ee
where the function $C_{E}(\tau,r)$ is nothing but the Euclidean 
Wilson loop, computed with an imaginary time coordinate $\tau$, 
with gauge fields periodic in $\tau\to \tau + \hbar/T$.

The expression that was obtained for $V_{>}(t,r)$ 
in ref.~\cite{static} reads 
(the superscript refers to the order in $g$; 
we keep $\hbar\neq 1$; 
and we assume the use of dimensional regularization) 
\ba
 V_{>}^{(2)}(t,r) 
 & = & 
 -\frac{g^2 C_F \hbar }{4\pi} \biggl[ 
 \mD + \frac{\exp(-\mD r)}{r}
 \biggr]  + 
 \delta V_{>}^{(2)}(t,r) 
 \;, \la{Vtr0}
 \\ 
 \delta V_{>}^{(2)}(t,r) 
 & = & 
 g^2 C_F \hbar
 \int \! \frac{{\rm d}^3\vec{p}}{(2\pi)^3}
  (1 - \cos p_3 r) 
  \times \nn 
 & \times & 
 \int_{-\infty}^{\infty} \! \frac{{\rm d} p^0}{\pi}
 p^0 
 \Bigl[ 
  e^{-i |p^0| t}
  + {\nB(|p^0|)} \Bigl( e^{-i |p^0| t} -  e^{i |p^0| t} \Bigr)  
 \Bigr] 
  \times \nn 
 & \times & 
 \biggl[ 
 \biggl( 
   \frac{1}{\vec{p}^2} - \frac{1}{(p^0)^2} 
 \biggr)  
 \rho_E(p^0,\vec{p}) 
 + 
 \biggl( 
  \frac{1}{p_3^2} - \frac{1}{\vec{p}^2}
 \biggr) 
 \rho_T(p^0,\vec{p}) 
 \biggr] 
 \;. \la{Vtr_new}
\ea
Here $C_F\equiv (\Nc^2-1)/2\Nc$;  
$\mD$ is the Debye mass parameter
(actually of dimensionality 1/distance rather than mass);
$\nB{}(x) \equiv 1/[\exp(\hbar x/T)-1]$ is 
the Bose distribution function;   
and we have chosen $\vec{r}\equiv (0,0,r)$.
The $r$-independent term in \eq\nr{Vtr0} amounts to twice 
a thermal mass correction for the heavy quark. 
For the gluon spectral functions $\rho_E, \rho_T$ we assume 
the conventions specified in appendix A of ref.~\cite{og2}. 

It can be observed that \eqs\nr{Vtr0}, 
\nr{Vtr_new} contain both a real and an imaginary part.
In particular, the familiar-looking structure in 
\eq\nr{Vtr0}, representing a Debye-screened Coulomb potential, 
is real, while the manifestly ``thermal'' part in \eq\nr{Vtr_new}, 
containing $\nB{}$, is 
purely imaginary. As pointed out in ref.~\cite{static}, this 
purely imaginary term remains non-zero in 
the limit $t\to\infty$, because of Bose-enhancement
at small frequencies, $\hbar|p^0|\ll T$ (cf.\ \eq\nr{limes}).  

The Bose-enhanced term corresponds to the physics 
of the {\em classical limit} $\hbar\to 0$, in which situation
$
 \nB(|p^0|) = T/\hbar |p^0|
$. 
In fact, all other terms vanish in this limit,
being multiplied by $\hbar$.
The classical potential then reads
\ba
 V_\rmi{cl}^{(2)}(t,r) 
 & = & 
 g^2 C_F T
 \int \! \frac{{\rm d}^3\vec{p}}{(2\pi)^3}
 (1 - \cos p_3 r) 
 \int_{-\infty}^{\infty} \! \frac{{\rm d} p^0}{\pi}
 \Bigl( e^{-i p^0 t} -  e^{i p^0 t} \Bigr)   
  \times \nn 
 & \times & 
 \lim_{\hbar \to 0} \biggl[ 
 \biggl( 
   \frac{1}{\vec{p}^2} - \frac{1}{(p^0)^2} 
 \biggr)  
 \rho_E(p^0,\vec{p}) 
 + 
 \biggl( 
  \frac{1}{p_3^2} - \frac{1}{\vec{p}^2}
 \biggr) 
 \rho_T(p^0,\vec{p}) 
 \biggr] 
 \;, \la{Vcl}
\ea
where we have also simplified the way in which $p^0$'s appear. 
For large times, 
\be
 \lim_{t\to\infty} \frac{e^{i p^0 t}-e^{-i p^0 t}}{p^0} 
 = 2 \pi i \; \delta(p^0)
 \;, \la{limes}
\ee
and we obtain
\ba
 V_\rmi{cl}^{(2)}(\infty,r) 
 & = & 
 2 i g^2 C_F T
 \int \! \frac{{\rm d}^3\vec{p}}{(2\pi)^3}
 ( 1 - \cos {p_3 r}) 
 \lim_{p^0\to 0} \lim_{\hbar \to 0}
 \frac{ \rho_E(p^0,\vec{p}) }{p^0} 
 \;. \la{Vasym_r}
\ea
If we also take the limit $r\to\infty$, 
the cosine-term in \eq\nr{Vasym_r} drops out. 
Assuming for a moment that the two limits and the integration
in \eq\nr{Vasym_r} commute,
and making use of the 
known Hard Thermal Loop form of $\rho_E(p^0,\vec{p})$
(\eq(B.13) of ref.~\cite{static}
shows $\rho_E$ at small $|p^0|$ with our conventions), 
containing the parameter $\mD^2$, 
then leads to the provisional result 
\be
 V_\rmi{cl}^{(2)}(\infty,\infty) 
 \stackrel{?}{=}
 \lim_{\hbar\to 0} 
 - i g^2 C_F T
 \int \! \frac{{\rm d}^3\vec{p}}{(2\pi)^3}
 \frac{\pi \mD^2}{|\vec{p}|(\vec{p}^2 + \mD^2)^2}
 = 
 \lim_{\hbar\to 0} 
 - i \frac{g^2 C_F T}{4\pi}
 = 
 - i \frac{g^2 C_F T}{4\pi}
 \;. 
 \la{Vasym} 
\ee
Note that in the quantum theory the same result is obtained
for the asymptotic value $\delta V^{(2)}_{>}(\infty,\infty)$~\cite{static}, 
and in this sense \eq\nr{Vasym} is indeed the correct physical expression. 

Now, given that the integral in \eq\nr{Vasym}
is finite, it might be assumed that the result is independent 
of the regularization procedure. 
It turns out that this argument is too naive:
in fact, $\mD^2$ diverges as $g^2 T^2/\hbar$ in dimensional 
regularization, indicating that the classical
limit may introduce ultraviolet singularities. 
In particular, if the ultraviolet is
regularized by a lattice rather than dimensionally, 
with a spatial lattice spacing $a$, 
then the limit becomes finite, 
$\lim_{\hbar\to 0} \mD^2 \propto g^2 T/a$~\cite{bms,cllattice}. 
Thus the classical limit in \eq\nr{Vasym_r} does exist, 
but the price to pay is that $\rho_E$ and subsequently 
$V_\rmi{cl}^{(2)}(\infty,\infty)$
depend on the details of the regularization procedure. 
In particular, carrying out the limits in the order
indicated by \eq\nr{Vasym_r} with lattice regularization, 
does not lead to the expression in \eq\nr{Vasym} 
(cf.\ \fig\ref{fig:Vasym} below). 

Fortunately, this problem is not too serious: an analogous
situation was met in studies of the sphaleron rate in the 
electroweak theory, yet 
classical lattice gauge theory simulations~\cite{oldclas} did yield 
non-perturbative physical information, once properly interpreted
(see, e.g., refs.~\cite{mr,bmr}). In our case, 
\fig\ref{fig:Vasym} implies 
that we cannot use classical lattice gauge 
theory simulations to compute corrections directly to 
\eq\nr{Vasym}. However, we can compute the analogues 
of \eqs\nr{Vcl}--\nr{Vasym} with Hard Thermal Loop perturbation
theory adapted to the ultraviolet physics of the classical 
lattice~\cite{bms,cllattice}, and compare subsequently
these results (the dashed curve in \fig\ref{fig:Vasym})
with a non-perturbative determination. 
In this way we can probe the 
{\em infrared sector} of thermal field theory, 
which indeed is classical in nature. 

%
\section{Perturbative real-time static potential 
in classical lattice gauge theory}
\la{se:clgt}

We assume that the theory is regularized by introducing 
a cubic spatial lattice, while the time 
coordinate is continuous. Gauge field configurations are generated 
with a Wilson-discretised Hamiltonian (cf.\ \eq\nr{Zcl} below), and evolved 
with the classical equations of motion (cf.\ \eqs\nr{dU}, \nr{dE} below).
The results depend on a single parameter, 
\be
 \beta \equiv \frac{2 C_A}{g^2 T a}
 \;, 
\ee
where $C_A \equiv \Nc$. 

The way to carry out perturbation theory in this situation
was worked out in refs.~\cite{bms,cllattice}. The procedure is analogous
to Hard Thermal Loop resummed perturbation theory~\cite{htl}, 
with technical differences originating from the different 
ultraviolet physics. We will refer to this procedure 
as Hard Classical Loop (HCL) perturbation theory. 

Let us start by introducing the notation
\be
 \tilde p_i \equiv \frac{2}{a} \sin \Bigl( \frac{a p_i}{2} \Bigr)
 \;, \quad
 \ring p_i \equiv \frac{1}{a} \sin(a p_i)
 \;, \quad
 \tilde \vec{p}^2 \equiv \sum_{i=1}^{3} \tilde p_i^2 
 \;, \quad
 \ring \vec{p}^2 \equiv \sum_{i=1}^{3} \ring p_i^2 
 \;.
 \la{latt_mom}
\ee
Also, the integration measure is denoted by 
\be
 \int \! {\rm d} \vec{p} \equiv 
 \int_{-\pi/a}^{\pi/a} \! 
 \frac{{\rm d}^3 \vec{p}}{(2\pi)^3}
 \;. \la{latt_int}
\ee
Then, we consider \eq\nr{Vcl}, with a few straightforward 
modifications following from the introduction of lattice regularization: 
\ba
 V_\rmi{cl}^{(2)}(t,r) 
 & = & 
 g^2 C_F T \!
 \int \! {\rm d} \vec{p} \, 
 ( 1 - \cos p_3 r ) 
 \int_{-\infty}^{\infty} \! \frac{{\rm d} p^0}{\pi}
 \Bigl( e^{-i p^0 t} -  e^{i p^0 t} \Bigr)   
  \times \nn 
 & \times & 
 \biggl[ 
 \biggl( 
   \frac{1}{\tilde \vec{p}^2} - \frac{1}{(p^0)^2} 
 \biggr)  
 \rho_E(p^0,\tilde \vec{p}) 
 + 
 \biggl( 
  \frac{1}{\tilde p_3^2} - \frac{1}{\tilde \vec{p}^2}
 \biggr) 
 \rho_T(p^0,\tilde \vec{p}) 
 \biggr] 
 \;. \la{Vcl_a}
\ea
Here, 
\be
 \rho_E(p^0,\tilde \vec{p}) \equiv \frac{1}{2 i} 
 \Bigl[ 
   \Delta_E(p^0 + i \epsilon,\tilde\vec{p}) - 
   \Delta_E(p^0 - i \epsilon,\tilde\vec{p})
 \Bigr]
 \;, \la{rhodef}
\ee
with $\epsilon = 0^+$, and the propagator $\Delta_E$ has the form
\be
 \Delta_E(p^0,\tilde\vec{p}) = 
 \frac{1}{\tilde \vec{p}^2 - (p^0)^2 + \Pi_E(p^0,\tilde \vec{p})}
 \;. \la{propdef}
\ee
The limit $\lim_{\hbar\to 0}$ is assumed everywhere but not shown 
explicitly. 

%
\subsection{Behaviour at finite times}

%
\begin{figure}[t]
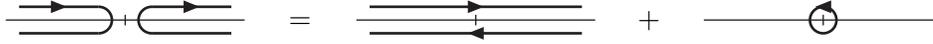


\begin{eqnarray*}
&& 
 \hspace*{-1cm}
 \ContA \quad = \quad 
 \ContB \quad + \quad 
 \ContC 
\end{eqnarray*}

\caption[a]{\small 
Integration contours for the classical real-time static potential. 
} 
\la{fig:contour}
\end{figure}
%

In order to simplify \eq\nr{Vcl_a}, it is convenient, 
following ref.~\cite{bl}, to view the $p^0$-integration 
as an integral in the complex plane, and to deform the contour suitably. 
As it stands, the integrand in \eq\nr{Vcl_a} is finite at $p^0 = 0$ 
(note that $\rho_E$ is linear in $p^0$ around the origin).  
However, the part multiplying $\rho_E$ contains a pole at $p^0 = 0$;
this pole just does not contribute because of the mentioned property
of $\rho_E$. Writing $\rho_E$ as in \eq\nr{rhodef}, this means 
that we can view the original integral as indicated in the left-most
drawing in \fig\ref{fig:contour}, and then also deform it accordingly. 
Subsequently, $\epsilon$ can be taken to be finite, because there are
no singularities outside of the real axis. Furthermore, assuming $t > 0$, 
terms multiplied by $\exp[i (p^0 + i \epsilon) t]$ and 
$\exp[-i(p^0 - i \epsilon) t]$ must vanish, 
because we can imagine taking $\epsilon$ arbitrarily large. 
The integrand can only decrease in this limit, and being 
multiplied by $\exp(-\epsilon t)$, the integral then vanishes
(it is a good cross-check of the numerics to verify the 
vanishing at any finite $\epsilon$). Finally, the symmetry properties of the 
integrand allow to reflect the lower of the remaining contours 
to the upper half-plane. We thus obtain
\ba
 & & \hspace*{-2cm}
 \im \biggl[ \frac{V_\rmi{cl}^{(2)}(t,r)}{g^2 T} \biggr]
 =  
 2 C_F 
 \int \! {\rm d} \vec{p} \, 
 ( 1 - \cos p_3 r ) 
 \biggl\{ 
 t\, \Delta_E(0,\tilde \vec{p}) - 
 \int_{-\infty}^{\infty} \! \frac{{\rm d} p^0}{2 \pi}
  e^{-i (p^0 + i \epsilon) t}   
 \times \nn 
 & \times & 
 \biggl[ 
 \biggl( 
   \frac{1}{\tilde \vec{p}^2} - \frac{1}{(p^0+i\epsilon)^2} 
 \biggr)  
 \Delta_E(p^0+i\epsilon,\tilde \vec{p}) 
 + 
 \biggl( 
  \frac{1}{\tilde p_3^2} - \frac{1}{\tilde \vec{p}^2}
 \biggr) 
 \Delta_T(p^0+i\epsilon,\tilde \vec{p}) 
 \biggr] 
 \biggr\}
 \;,  
 \la{Vcl_contour}
\ea
where the first term is the contribution of the pole in \fig\ref{fig:contour}.

Let us stress that the integration 
in \eq\nr{Vcl_contour} is independent of the value of $\epsilon > 0$, 
since there are no poles in the upper half-plane. Checking the independence
in practice offers another cross-check for the accuracy of the numerical
integration. Naturally, small values of $\epsilon$ are difficult, 
because the integrand becomes strongly peaked around the origin, 
while large values of $\epsilon$ are also difficult, because
the latter term is multiplied by $\exp(\epsilon t)$, whereby
the numerical errors of the integration are exponentially
amplified at large $t$; a useful compromise appears to be 
$\epsilon \simeq 1/a$. In general, 
it is advantageous to decrease $\epsilon$ 
when increasing $t$.

In order to insert the propagators $\Delta_E, \Delta_T$, we need 
to know the self-energies $\Pi_E$, $\Pi_T$ (cf.\ \eq\nr{propdef}). 
Starting from the spatial part of 
the gluon self-energy~\cite{bms,cllattice}
\be
 \Pi_{ij}(p^0,\tilde \vec{p}) =  2 g^2 T C_A 
   \int \! {\rm d}\vec{q} \, 
   \frac{1}{\tilde \vec{q}^2}
   \frac{p^0 v_i v_j }{p^0 - \tilde \vec{p} \cdot \vec{v}} 
  \;, 
\ee
with 
\be
 v_i \equiv \frac{\ring q_i}{\sqrt{\tilde \vec{q}^2}}
 \;, 
\ee
and employing the projection operators $P^E_{\mu\nu}, P^T_{\mu\nu}$
defining $\Pi_E, \Pi_T$ (we use the conventions
specified in appendix~B of ref.~\cite{static}), 
we obtain 
\ba
 \Pi_E(p^0,\tilde \vec{p})
 \!\! & = & \!\! 
 2 g^2 T C_A
 \biggl( 
  1 - \frac{(p^0)^2}{\tilde \vec{p}^2}
 \biggr)
 \biggl(
   \frac{\Sigma}{4\pi a}
   -
   \int \! {\rm d}\vec{q} \, 
   \frac{1}{\tilde \vec{q}^2}
   \frac{p^0}{p^0 - \tilde \vec{p} \cdot \vec{v}} 
 \biggr)
 \;, \la{PiE} \\ 
  \Pi_T(p^0,\tilde \vec{p})
 \!\! & = & \!\!
 g^2 T C_A
 \biggl[ 
  \frac{(p^0)^2}{\tilde \vec{p}^2}
  \biggl(
   \frac{\Sigma}{4\pi a}
   -
   \int \! {\rm d}\vec{q} \, 
   \frac{1}{\tilde \vec{q}^2}
   \frac{p^0}{p^0 - \tilde \vec{p} \cdot \vec{v}} 
 \biggr)
 + 
   \int \! {\rm d}\vec{q} \, 
   \frac{\ring \vec{q}^2}{(\tilde \vec{q}^2)^2}
   \frac{p^0}{p^0 - \tilde \vec{p} \cdot \vec{v}} 
 \biggr]
 \;. \la{PiT} \hspace*{1.0cm}
\ea
Here 
\be
  \frac{\Sigma}{4\pi a}
  \equiv 
  \int\!{\rm d}\vec{q} \, \frac{1}{\tilde \vec{q}^2} 
  \;,
  \la{Sigma}
\ee
where $\Sigma\approx 3.175911535625$  is a trigonometric factor which 
can be expressed in terms of the complete elliptic integral of 
the first kind~\cite{framework}.
Note that with the form in \eq\nr{PiE}, the combination
containing $\Delta_E$ in \eq\nr{Vcl_contour} becomes
\be
 \biggl( 
   \frac{1}{\tilde \vec{p}^2} - \frac{1}{(p^0+i\epsilon)^2} 
 \biggr)  
 \Delta_E(p^0+i\epsilon,\tilde \vec{p}) 
 = 
 - \frac{1}{(p^0+i\epsilon)^2} 
 \frac{1}
 {\tilde \vec{p}^2 + 
 2 g^2 T C_A
 \Bigl(
   \frac{\Sigma}{4\pi a}
   -
   \int \! {\rm d}\vec{q} \, 
   \frac{1}{\tilde \vec{q}^2}
   \frac{p^0}{p^0 - \tilde \vec{p} \cdot \vec{v}} 
 \Bigr)
 } 
 \;. 
\ee

Denoting the square brackets in \eq\nr{Vcl_contour} by 
$\mathcal{I}(p^0 + i\epsilon,\tilde \vec{p})$, and making 
use of the properties
\be
 \mathcal{I}(- p^0 + i\epsilon,\tilde \vec{p})
 = \mathcal{I}(p^0 - i\epsilon,\tilde \vec{p})
 = \Bigl[ \mathcal{I}(p^0 + i\epsilon,\tilde \vec{p}) \Bigr]^*
 \;, 
\ee
the integral over $p^0$ can furthermore be reduced to regular 
cosine and sine transforms: 
\ba
 & & \hspace*{-1.5cm}
 \int_{-\infty}^{\infty} \! \frac{{\rm d} p^0}{2 \pi}
 e^{-i (p^0 + i \epsilon) t}   
 \mathcal{I}(p^0 + i\epsilon,\tilde \vec{p})
 \nn 
 & & = 
 \frac{e^{\epsilon t}}{\pi}
 \int_0^\infty\! {\rm d} p^0 \, 
 \Bigl\{
  \cos(p^0 t)
  \re \Bigl[
  \mathcal{I}(p^0 + i\epsilon,\tilde \vec{p})
  \Bigr] 
  + 
  \sin(p^0 t)
  \im \Bigl[ 
  \mathcal{I}(p^0 + i\epsilon,\tilde \vec{p})
  \Bigr] 
 \Bigr\}
  \;. 
\ea
Though efficient routines for such transforms exist, it is also
clear that the accuracy requirements grow exponentially with $t$, 
so that very large times are difficult to reach. 

In a practical lattice study, the system possesses not only 
a finite lattice spacing, but also a finite extent, $L = N a$, 
where $N$ is the number of lattice points. We assume that the 
box is cubic and that periodic boundary conditions are imposed
in every direction. Furthermore, let us assume that we use 
changes of integration variables to write the momentum 
integrations over the ``positive'' octant only, 
\be
 \int_{-\pi/a}^{\pi/a} \! \frac{{\rm d}p_i}{2\pi} \mathcal{F}(p_i)
 =  \int_{0}^{\pi/a} \! \frac{{\rm d}p_i}{2\pi} \mathcal{G}(p_i)
  \;, \quad \mathcal{G}(p_i) \equiv 
  \Bigl[ \mathcal{F}(p_i) + \mathcal{F}(-p_i) \Bigr]
  \;, \quad i=1,2,3
  \;.
\ee
In a finite volume this then goes over into
\be
 \int_{0}^{\pi/a} \! \frac{{\rm d}p_i}{2\pi} \mathcal{G}(p_i)
 \longrightarrow
 \frac{1}{Na}
 \biggl[ 
  \fr12 
  \mathcal{G}( 0 )
  + 
  \sum_{i=1}^{N/2 - 1}
  \mathcal{G}\Bigl( \frac{2 \pi i}{N a} \Bigr)
  + 
  \fr12 
  \mathcal{G}\Bigl( \frac{\pi}{a} \Bigr)
 \biggr]
 \;. \la{sum}
\ee
Note that in finite volume, the analytically known integral 
in \eq\nr{Sigma} should also be replaced by a numerically
evaluated sum.

Now, \eq\nr{sum} contains also a contribution from the 
zero-mode, $p_i = 0$. Its treatment requires in general some care. 
In $\Pi_E, \Pi_T$, loop momenta are by definition ``hard'': it is 
sensible (and in fact necessary) to leave out the zero-mode. 
In the remaining sum in \eq\nr{Vcl_contour}, 
in contrast, momenta could be soft: we thus keep 
the contribution of the zero-mode as well (even though 
the practical effect is small). Note that for the zero-mode, 
\be
 \Pi_E(p^0,\vec{0}) = \Pi_T(p^0,\vec{0}) = 
 \omega_\rmi{pl}^2
 \;, \quad
 \omega_\rmi{pl}^2 \equiv 
 \fr23 g^2 T C_A \int \! {\rm d}\vec{q} \, 
 \frac{\ring \vec{q}^2}{(\tilde \vec{q}^2)^2}
 \;, 
\ee
where the integration can be replaced by a sum (without zero-mode)
as before.\footnote{%
  In infinite volume, 
  $
    \omega_\rmi{pl}^2 = g^2 T C_A (3\Sigma/2\pi - 1)/6a 
  $, 
  where $\Sigma$ is the constant in \eq\nr{Sigma}~\cite{bl}.
  } 
Then the combination in \eq\nr{Vcl_contour} becomes
\ba
 & & \hspace*{-1.2cm}
 (1 - \cos p_3 r)
 \biggl[ 
 \biggl( 
   \frac{1}{\tilde \vec{p}^2} - \frac{1}{(p^0+i\epsilon)^2} 
 \biggr)  
 \frac{1}{\tilde \vec{p}^2 - (p^0+i\epsilon)^2 + \omega_\rmi{pl}^2}
 + 
 \biggl( 
  \frac{1}{\tilde p_3^2} - \frac{1}{\tilde \vec{p}^2}
 \biggr) 
 \frac{1}{\tilde \vec{p}^2 - (p^0+i\epsilon)^2 + \omega_\rmi{pl}^2}
 \biggr]  
 \nn 
 & \stackrel{\vec{p}=\vec{0}}{\longrightarrow} & 
 \frac{r^2}{2} \frac{1}{ - (p^0+i\epsilon)^2 + \omega_\rmi{pl}^2}
 \;.   
\ea

\begin{figure}[t]


\centerline{%
\epsfysize=9.0cm\epsfbox{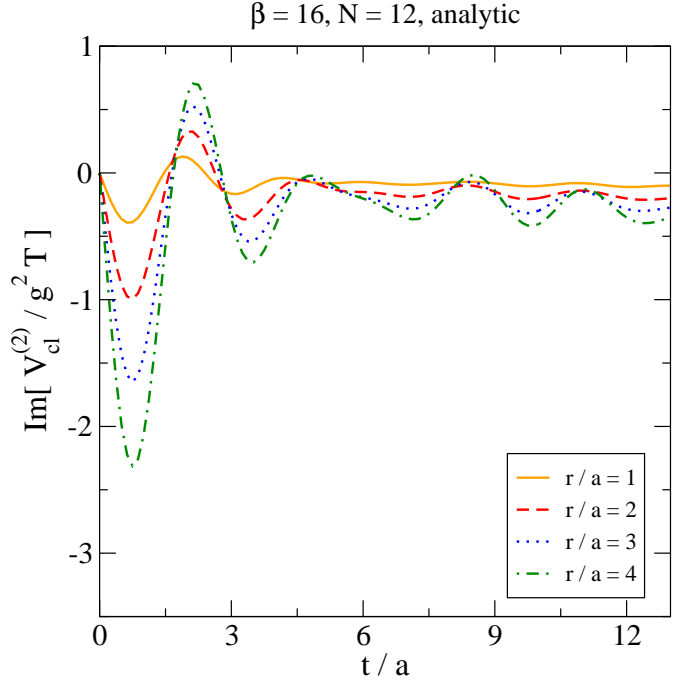}%
}

\caption[a]{\small 
The imaginary part of the classical real-time static potential,
to leading non-trivial order in HCL-resummed perturbation theory
(\eq\nr{Vcl_contour} with $\int\!{\rm d}\vec{p}$ replaced by 
a finite-volume sum), for $\beta=16$, $N=12$, $\Nc = 3$. 
} 
\la{fig:ImV}
\end{figure}

In practice, 
for the values $N\ge 12$ that we have used,
finite-volume effects are almost invisible at small times. 
Perturbative finite-volume effects grow rapidly with time, 
however, and also with distance. On the other hand, perturbation theory
tends to overestimate their significance, since it lacks the mass
gap generated by the confining dynamics. At the same time, 
whenever justified, it appears to be numerically advantageous 
to use the finite-volume expressions, which contain a six-fold exact 
sum, rather than to approximate the corresponding infinite-volume continuous
six-dimensional momentum integration numerically. Therefore we plot 
the perturbative expression only in the range where the perturbative
finite-volume effects are small, $t/a\lsim 10$; 
an example of a result is shown in \fig\ref{fig:ImV}. 

%
\subsection{Value in the large-time limit}
 
As mentioned, it is not easy to evaluate numerically the HCL-resummed
perturbative expression in \eq\nr{Vcl_contour}, 
once the time coordinate becomes large: 
$\epsilon$ should be decreased, whereby the integrand becomes 
strongly peaked; and one should replace the finite-volume sums
with infinite-volume momentum integrals, whereby the numerical
cost increases. To get a handle on this limit 
we can, however, proceed in another way, 
without making use of the contour trick, 
and thereby obtain the correct version of \eq\nr{Vasym} on 
an infinite spatial lattice. The starting point is then \eq\nr{Vasym_r}.

For infinitesimally small $\epsilon$, the small-$p^0$ behaviour
of $\Pi_E$ reads (cf.\ \eq\nr{PiE})
\be
 \Pi_E(p^0+i 0^+,\tilde \vec{p})
 = 
 2 g^2 C_A T 
 \biggl[ 
  \frac{\Sigma}{4\pi a}
  + i \pi p^0 
  \int\!{\rm d}\vec{q} \, \frac{1}{\tilde \vec{q}^2}
  \delta 
  \Bigl( 
    \frac{\tilde \vec{p} \cdot \ring \vec{q}}{\sqrt{\tilde \vec{q}^2}}
  \Bigr) 
 \biggr]
  + \rmO((p^0)^2) 
 \;.
\ee
Making use of 
the definition in \eq\nr{rhodef}, 
the asymptotic value from \eq\nr{Vasym_r} can be written as
\ba
 \im \biggl[ \frac{V^{(2)}_\rmi{cl}(\infty,r)}{g^2 T} \biggr]
 = 
 -\frac{\pi C_F C_A^2}{\beta}
  \int_{0}^{1} \! {\rm d}^3 \vec{x} \, 
  \frac{1 - \cos(\pi x_3 r/a)}{(\tilde \vec{x}^2 + C_A^2 \Sigma/\pi\beta)^2}
  \int_{-1}^{1}  \! {\rm d}^3 \vec{y} \,
  \frac{\delta(\tilde \vec{x} \cdot \ring \vec{y})}{(\tilde \vec{y}^2)^{1/2}}
 \;, 
\ea
where we have gone over to  a notation
where the integration variables are made dimensionless
by going to lattice units, and the integration range
is restricted to the unit box and its reflections:
\be
 \tilde x_i \equiv {2} \sin \Bigl( \frac{\pi x_i}{2} \Bigr)
 \;, \quad
 \ring x_i \equiv \sin(\pi x_i)
 \;, \quad 
 x_i \in (-1,1) 
 \;. \la{latt_mom_x}
\ee
Moreover, we have made use of the symmetry of the integrand, 
in order to restrict the integration to positive $x_i$.

\begin{figure}[t]


\centerline{%
\epsfysize=9.0cm\epsfbox{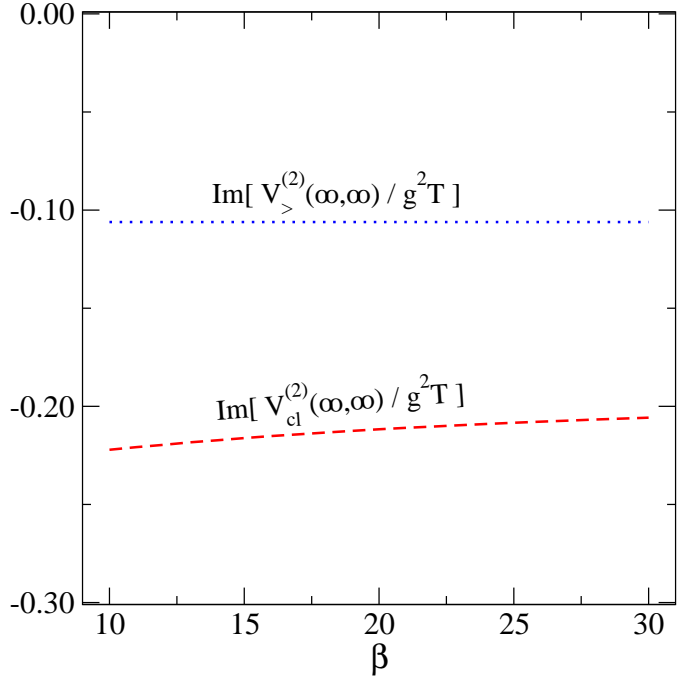}%
}

\caption[a]{\small 
The asymptotic value of the real-time static potential $V_\rmi{cl}(t,r)$, 
on an infinite lattice, to leading non-trivial order in 
HCL-resummed perturbation theory (\eq\nr{Vcl_asym}), for $\Nc = 3$. 
For comparison, we also show the expression on the right-hand
side of \eq\nr{Vasym}, which corresponds to the asymptotic value 
of the potential $V^{(2)}_{>}(t,r)$ in the full continuum quantum theory. 
} 
\la{fig:Vasym}
\end{figure}

Among the eight octants of the $\vec{y}$-integration, 
the $\delta$-function gets realised in six only, and by changes
of integration variables we can combine all the contributions together: 
\ba
 \im \biggl[ \frac{V^{(2)}_\rmi{cl}(\infty,r)}{g^2 T} \biggr]
 & = &
 -\frac{2 \pi C_F C_A^2}{\beta}
  \int_{0}^{1} \! {\rm d}^3 \vec{x} \,
  \frac{3 - \cos(\pi x_1 r/a) - \cos(\pi x_2 r/a) - \cos(\pi x_3 r/a)}
  {(\tilde \vec{x}^2 + C_A^2 \Sigma/\pi\beta)^2}
  \times \nn & \times & 
  \int_{0}^{1} \! {\rm d}^3 \vec{y} \,
  \frac{\delta(
  \tilde x_1 \ring y_1 + 
  \tilde x_2 \ring y_2 - 
  \tilde x_3 \ring y_3)}
  {
   (\tilde \vec{y}^2)^{1/2}}
  \;.  \la{Vcl_asym}
\ea
It is now straightforward to carry out the integration over, say, $x_3$,
to remove the $\delta$-function, 
and also to make use of the symmetry of the remaining integrand in 
$x_1 \leftrightarrow x_2$. The 5-dimensional integral left over 
can be evaluated numerically without too much trouble. 
The result is shown in \fig\ref{fig:Vasym}, for $\Nc = 3$ and the 
case $r/a=\infty$, when the cosine-term does not contribute.
Values of \eq\nr{Vcl_asym} at finite $r/a$ can be found in Table~1 below. 
Comparing Table~1 with \fig\ref{fig:Vasym}, we observe that for, say, 
$\beta = 16$, the distance $r/a = 4$ gives a value which is 
already relatively close to the asymptotic one. To summarise, 
in resummed perturbation theory 
$
 \im [ {V^{(2)}_\rmi{cl}(\infty,r)} 
     ]
$
is definitely non-zero at all $r\neq 0$.

%
\section{Classical lattice gauge theory simulations}
\la{se:sim}

The computation in the previous section was based on resummed
perturbation theory, but it was only carried out to a fixed order. 
Let us try to estimate the expansion parameters of such a computation. 
Using continuum notation, the vertices of each new loop order
bring in a factor $g^2T$. At the same time, the mass scales appearing
in the dynamics are the ultraviolet cutoff scale, $\Lambda \sim 1/a$, 
as well as the confinement scale of three-dimensional Yang-Mills
theory, $\mG \sim g^2 T$~\cite{linde,gpy}. The resummation accounts
for the dominant influence of the hard cutoff scale $\Lambda$ on the 
dynamics of the infrared modes, and is associated with corrections
of the type $g^2 T/\Lambda \sim 1/\beta$. However, it does not 
account for the self-interactions of the infrared modes, which may 
lead to an expansion parameter of the type $g^2 T/\mG \sim 1$.  
Therefore, we would like to compare the resummed perturbative
result with a non-perturbative numerical computation. 

The practical procedure of generating classical gauge field 
configurations is the following~\cite{oldclas}. Since our observable
will be gauge-invariant, we may choose a gauge;  it is convenient 
to work in the temporal gauge, $U_t(\vec{x},t)=\unit$. 
Here $U_\mu$ is an SU($\Nc$) link matrix. 
The canonical degrees of freedom are the 
spatial link variables $U_i(\vec{x},t)$ and the matrix-valued
``canonical momenta'' $E_i(\vec{x},t)$, which transform in the 
adjoint representation. We denote the generators of the gauge group by 
$T^a$, and assume them normalised through $\tr[T^aT^b] = \delta^{ab}/2$. 
Furthermore, $E_i \equiv E_i^a T^a$, 
$S_{ij}(\vec{x}) \equiv U_j(\vec{x})U_i(\vec{x}+j)U^\dagger_j(\vec{x}+i)$,
where $i \equiv a \hat e_i$, 
and $U_{-i}(\vec{x}) \equiv U_i^\dagger(\vec{x} - i)$. 

With this notation, the procedure starts by generating initial 
configurations (at time $t=0$) according to the partition 
function~\cite{oldclas} 
\be
 \mathcal{Z} = 
 \int\!\mathcal{D} U_i 
 \mathcal{D}E_i \,
 \delta(G)
 \exp\biggl\{-\beta \sum_\vec{x} 
  \biggl[ \sum_{i< j} 
     \Bigl( 1 - \frac{1}{C_A} \re \tr P_{ij} \Bigr) 
    +\sum_{i} \tr (E_i^2) \biggr] \biggr\}
 \;,  \la{Zcl}
\ee
where $P_{ij}$ is the spatial plaquette, and the Gauss law function reads
\be
 G(\vec{x},t) = \sum_i \Bigl[ 
 E_i(\vec{x},t) - U_{-i}(\vec{x},t) E_i(\vec{x}-i,t)
 U_{-i}^\dagger(\vec{x},t)
 \Bigr]
 \;.  \la{Gauss}
\ee 
To obtain configurations extending to $t > 0$, 
we solve the equations of motion
\ba
 a\, \partial_t U_i(\vec{x},t) & = & i \, (2 C_A)^\fr12 
 E_i(\vec{x},t) U_i(\vec{x},t) 
 \;, \la{dU} \\
 a\, \partial_t E^b_i(\vec{x},t) & = & 
 - \left(\frac{2}{C_A}\right)^\fr12 \im \tr \Bigl[ 
 T^b U_i(\vec{x},t) \sum_{|j|\neq i} S^\dagger_{ij}(\vec{x},t)
 \Bigr] 
 \;. \la{dE}
\ea
These four-dimensional configurations are then used for 
evaluating the real-time observables. 
In all that follows, we fix $C_A = \Nc = 3$, even though
we have also carried out some simulations at $\Nc = 2$  as a crosscheck.

It is worth stressing that in \eqs\nr{dU}, \nr{dE}, 
the lattice spacing is finite in spatial directions only. 
In practice, of course, the time direction needs to be discretised
as well, but with a very small lattice spacing, $a_t \ll a$.
As a check of the time evolution, it is useful to control the  
conservation of the Gauss law and of the total energy.

To specify the observable to measure, we adopt the definition
in \eq\nr{Seq2} as our non-perturbative starting point. 
The object appearing here is a specific analytic continuation
of the Euclidean Wilson loop, and corresponds 
formally to a time ordering generally denoted 
with the subscript $(...)_{>}$~\cite{books}:
\be
 C_{>}(t,r) \equiv C_E(it,r)
 \;. 
\ee
At the same time, the classical ($\hbar\to 0$) part of 
the analytic continuation, which we denote by $C_\rmi{cl}(t,r)$,
is independent of time ordering
($
 \lim_{\hbar\to 0} C_{>} = 
 \lim_{\hbar\to 0} C_{<} = 
 \lim_{\hbar\to 0} [C_{>} + C_{<}]/2 
$). 
In fact, $C_\rmi{cl}(t,r)$
is nothing but the classical Wilson loop, defined in 
Minkowski time. (Note that having chosen the gauge
$U_t = \unit$, the classical Wilson loop amounts really 
to a two-point correlation function of two spatial Wilson 
lines, both of which are local in time.) 
The classical static potential is then measured from 
\be
 i \partial_t C_\rmi{cl}(t,r) \equiv V_\rmi{cl}(t,r) C_\rmi{cl}(t,r)
 \;.  \la{Seq3}
\ee
It turns out that $C_\rmi{cl}(t,r)$ is real for all times
(within statistical errors), and slowly decaying. Therefore, 
$V_\rmi{cl}(t,r)$ is purely imaginary, with a negative imaginary part. 

The technical implementation of our simulation follows 
earlier work~\cite{oldclas,mr}. However, to speed up thermalization, 
we have implemented the idea mentioned in ref.~\cite{gdm}, whereby
the link variables $U_i$ are first pre-thermalized with regular
Monte Carlo techniques in the dimensionally reduced SU(3) + adjoint
Higgs theory (we use the code described in ref.~\cite{ah}).
Since it is non-trivial to match the parameters of that theory
and our effective theory  
exactly, those configurations are not yet fully thermalized.
However, this is not a problem, 
they now need only to be 
evolved for a short time \`a la refs.~\cite{oldclas,mr}, 
in order to reach the correctly thermalized configurations
corresponding to the exact parameters of \eq\nr{Zcl}.

We have carried out simulations mostly with 
$\beta = 16$; since analytic HCL predictions
also refer to a finite value of $\beta$, there is no need to 
carry out a continuum extrapolation (cf.\ \fig\ref{fig:Vasym}).  
As typical lattice extents we have used $N = 12$ and $N=16$; 
the difference of the results between these two is only at the percent 
level (cf.\ Table 1 below). 
The time variable is discretised with a spacing $a_t$, 
with a value $a_t/a = 0.01$; measurements are recorded
every 10th time step. We stress that thermalization is only
carried out in the beginning, while the subsequent time evolution
is deterministic and follows \eqs\nr{dU}, \nr{dE}.

\begin{figure}[t]


\centerline{%
\epsfysize=9.0cm\epsfbox{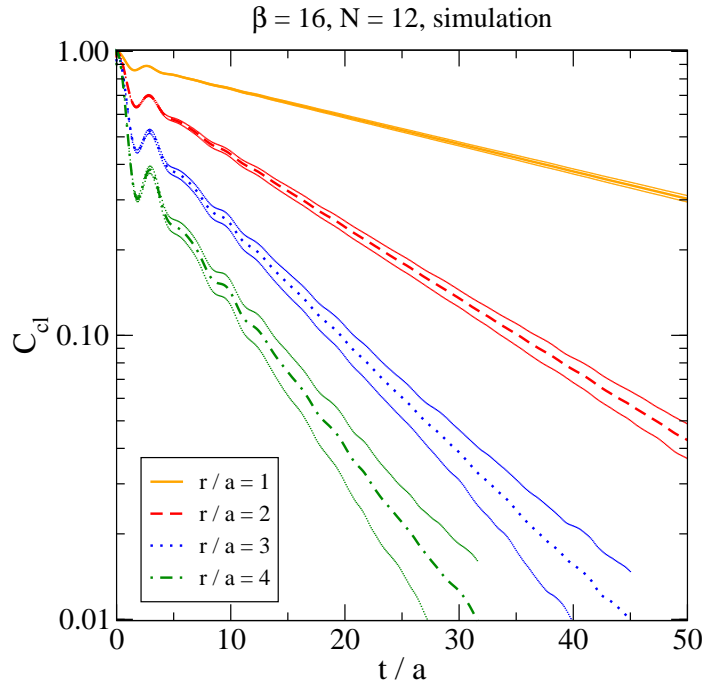}%
}

\caption[a]{\small 
The classical Wilson loop, measured with classical 
lattice gauge theory simulations, as a function of time in units
of the spatial lattice spacing, for $\beta = 16, N = 12, \Nc = 3$. 
} 
\la{fig:Ccl_simu}
\end{figure}

\begin{figure}[t]


\centerline{%
\epsfysize=9.0cm\epsfbox{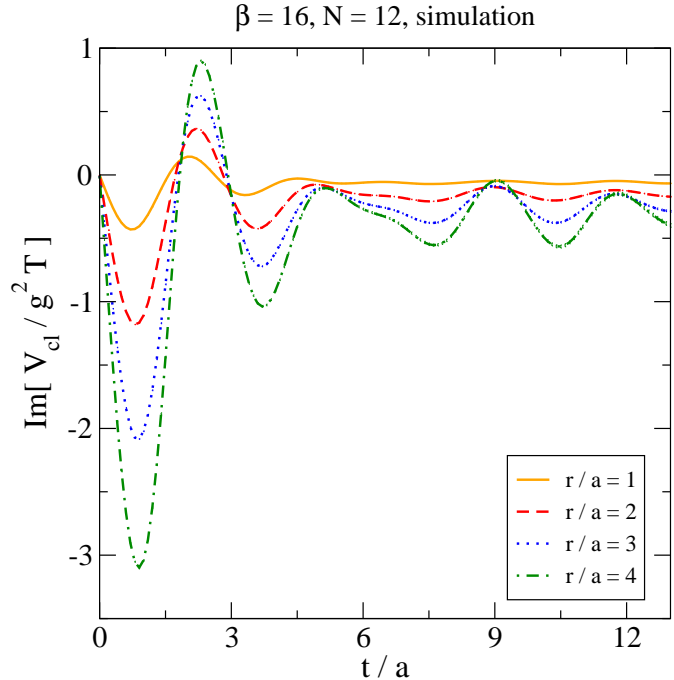}%
}

\caption[a]{\small 
The imaginary part of the real-time static potential, 
measured with classical lattice gauge theory simulations, 
as a function of time in spatial lattice units
(with the same parameter values as in \fig\ref{fig:Ccl_simu}). 
Vertical lines indicate statistical errors, 
but they are almost invisible in this time range. 
} 
\la{fig:ImV_simu}
\end{figure}

A representative result for the classical Wilson loop is shown 
in \fig\ref{fig:Ccl_simu}. The corresponding potential, 
extracted from \eq\nr{Seq3}, is shown in \fig\ref{fig:ImV_simu}.
The result can be compared with \fig\ref{fig:ImV}, showing
the HCL-resummed perturbative prediction with the same 
parameter values. The general shapes are seen to match
each other to a remarkable degree. On closer inspection, 
however, the amplitude of the oscillations is larger in the 
simulation; the frequency of oscillations is smaller (the oscillation period 
is larger); and the absolute value of the potential is larger
(the imaginary part is more negative).

\begin{table}[t]
\begin{center}
\begin{tabular}{|c|c|c|c|c|l|l|l|l|}
\hline
\multicolumn{5}{|c|}{parameters} & 
\multicolumn{4}{|c|}{value of $\im[V_\rmi{cl}(\infty,r)/g^2T]$} \\ \hline
$\beta$ & $N$ & $a \mD^\rmii{(bare)}$ & method & confs &    
$r/a = 1$   &  $r/a = 2$ & $r/a = 3$ & $r/a = 4$ \\
\hline\hline
$16.0$ & $12$ & $0.00$ & simulation & 200
 &  -0.060(2)
 &  -0.156(8)
 &  -0.246(26)
 &  -0.319(56) \\
\hline
$16.0$ & $16$ & $0.00$ & simulation & 160
 &  -0.059(2) 
 &  -0.155(8)~
 &  -0.245(22)
 &  -0.326(48) \\
\hline
$16.0$ & $12$ & $0.21$ & simulation & 200
 &  -0.059(2)
 &  -0.147(7)~
 &  -0.229(23)
 &  -0.297(51) \\
\hline
$16.0$ & $12$ & $0.35$ & simulation & 182
 &  -0.030(2) 
 &  -0.064(5)~
 &  -0.096(12)
 &  -0.118(21) \\
\hline
$13.5$ & $12$ & $0.25$ & simulation & 142
 &  -0.071(2) 
 &  -0.174(10)
 &  -0.270(33)
 &  -0.341(97) \\
\hline\hline
$16.0$ & $\infty$ & $0.00$ & analytic & ---
 &  -0.0601
 &  -0.1145
 &  -0.1507
 &  -0.1737 \\
\hline
\end{tabular}
\vskip 0.5cm
\caption{\small
The asymptotic values $\im[V_\rmi{cl}(\infty,r)/g^2T]$, 
obtained by fitting a constant to data in the range 
$t/a = 15 ... 30$. The numbers in parentheses indicate
the uncertainties of the last digits. The cases $a \mD^\rmii{(bare)} = 0.00$
refer to the classical theory without HTL degrees of freedom. The 
bottom row gives the perturbative values from \eq\nr{Vcl_asym};
the perturbative result at $r/a=\infty$ is -0.2152.
}
\label{T1}
\end{center}
\end{table}

In order to quantify the difference, we note that 
at large times, the potential obtains a constant value
(or, in terms of the Wilson loop, $C_\rmi{cl}$ decays
exponentially, cf.\ \fig\ref{fig:Ccl_simu}). 
We estimate this value by fitting a constant to data 
in the range $t/a = 15 ... 30$, where initial transients
have died out, yet the statistical errors are still relatively 
small for all parameter values that we have used. The results of 
the fits are shown in Table~1. Unfortunately, statistical
errors rapidly increase with $r/a$, and we are not able
to go to large enough values for the $r$-dependence to have flattened off. 
Nevertheless, the values at $r/a = 4$ already indicate that the
asymptotic value is larger (in absolute magnitude) 
than the analytic HCL estimate, 
by some 50 -- 100\%. The fact that there thus appears
to be somewhat more ``damping'' in the non-perturbative
classical dynamics than in the HCL estimate
is not surprising: other observables have 
yielded indications of a similar pattern~\cite{bl}.

We have also carried out so-called HTL simulations, 
both with the implementation based on treating the 
velocities of the hard particle degrees of freedom
with spherical harmonics~\cite{bmr,hr}, and through 
a discretization based on platonic solids~\cite{rrs}. 
The HTL-simulations introduce a new
parameter, which we refer to as $\mD^\rmii{(bare)}$. 
For small $\mD^\rmii{(bare)}$, say $a \mD^\rmii{(bare)} \lsim 0.2$, 
the results are practically identical with those of the classical
simulations (cf.\ Table~1). With increasing 
$\mD^\rmii{(bare)}$, say $a \mD^\rmii{(bare)} \gsim 0.35$, 
we see some discrepancies; in particular, the asymptotic value 
$\im[V_\rmi{cl}(\infty,r)]$ decreases in absolute magnitude 
(cf.\ Table~1), as one would 
expect in a situation where a lattice-induced dynamical Debye 
screening is overtaken by a continuum-like parameter
(cf.\ \fig\ref{fig:Vasym}). At the same
time, this method is not really bringing us closer to 
the {\em physical} continuum limit (i.e.\ the continuum
limit of the quantum theory), since in that situation
the bare parameter $[\mD^\rmii{(bare)}]^2$ should in fact become
negative as $\beta$ is increased, in order to cancel ultraviolet
divergences from the dynamics. Unfortunately, the 
implementation of HTL-simulations that we have followed, 
based on refs.~\cite{bmr,rrs}, does not allow to simulate
at $[\mD^\rmii{(bare)}]^2 < 0$, and a single bare parameter would
in any case not allow to renormalise all the observables that can 
be measured with classical lattice gauge theory~\cite{bms,cllattice}.
Therefore, we omit a more detailed
discussion of the HTL-simulations from here.

%
\section{Conclusions}
\la{se:concl}

The purpose of this paper has been to elaborate on the fact 
that the finite-temperature real-time static potential extracted
from an analytic continuation of the Euclidean Wilson loop, which 
can (at least to some order in perturbation theory)
be inserted into a Schr\"odinger-type equation governing
the behaviour of a certain heavy quarkonium Green's function, 
contains an imaginary part. As discussed elsewhere~\cite{og2}, 
this imaginary part has an impact on the heavy quarkonium 
spectral function at temperatures above a few hundred MeV, 
introducing a width to (the tip of) the resonance peak.  

Physically, the imaginary part implies that quarkonium at high temperatures
should not be thought of as a stationary state. Rather, the norm of its wave
function decays exponentially with time. 

The imaginary part emerges from Bose-enhanced infrared dynamics 
and, in field-theory language, is classical in nature 
(in particle language, it corresponds to a net disappearance 
of low-energy off-shell gauge particles,  
due to inelastic 
$
 2\to 1
$
and
$
 1\to 2
$
scatterings with the hard particles
in the plasma). 
We have computed the imaginary part with Hard Classical Loop
resummed perturbation theory, 
and with non-perturbative classical lattice gauge theory 
simulations. The comparison of the results, \fig\ref{fig:ImV}
and \fig5, or Table 1, shows reasonable qualitative agreement.
We conclude that non-perturbative colour-magnetic fields do 
{\em not} play a dominant role for the imaginary part of 
the real-time static potential; however, 
the non-perturbative corrections, together with higher-order
perturbative terms, are important on the quantitative level, 
bringing about some 50 -- 100\% increase in the absolute value
of the imaginary part at large times
(for $\beta = 16$), akin in magnitude to the correction
observed for the static Debye screening mass~\cite{ah}. In any case, 
our study confirms that an imaginary part exists, 
and suggests that Hard Loop perturbation theory presumably
provides for a reasonable first estimate for it 
also in the full quantum theory. 

For physical applications,
such as determining the quarkonium spectral function,  
it is essential to use the full quantum theory, rather
than the classical one. 
Moreover, it is  convenient to use 
dimensional regularization. Finally, as argued in ref.~\cite{og2}, 
the static potential should be evaluated at $t\gg r$. 
The perturbative static potential in this limit, 
$V^{(2)}_{>}(\infty,r)$, 
including both a real and an imaginary part, can be found in 
Eqs.~(4.3), (4.4) of ref.~\cite{static}, and has already been 
employed for estimating the quarkonium spectral function 
in ref.~\cite{og2}. Increasing the imaginary part by some 
50 -- 100\% in the results of ref.~\cite{og2} lowers and widens 
the quarkonium peak, but the effect is not dramatic; 
in general, it appears that the spectral function is more sensitive to 
the real than the imaginary part of the real-time static potential.

As the next step of our program, 
we would therefore like to get a non-perturbative
handle also on the real part of the real-time static potential, 
$V_{>}(t,r)$, entering the Schr\"odinger-equation. In particular, 
it would be important to clarify its connection to the other static 
potentials that are being used for studying the spectral function 
of heavy quarkonium in high-temperature QCD 
(for recent work and references see, e.g., ref.~\cite{mp}).

%
\section*{Acknowledgements}

We wish to thank P.~Romatschke for fruitful 
collaboration during initial stages of this investigation. 
M.L.\ acknowledges useful discussions with Seyong Kim.
Our work was partially supported by the BMBF project
{\em Hot Nuclear Matter from Heavy Ion Collisions 
     and its Understanding from QCD}.


\end{document}